\documentclass[preprint,eqsecnum,nofootinbib]{revtex4}
\usepackage{amssymb,amsbsy,bm,amsmath,psfrag,dsfont}
\usepackage{hyperref}
\usepackage{graphicx}
\usepackage{epsfig}

\begin{document}

\newcommand{\ba}{\begin{array}}
\newcommand{\ea}{\end{array}}
\newcommand{\nn}{\nonumber}
\newcommand{\no}{\noindent}
\newcommand{\la}{\lambda}
\newcommand{\si}{\sigma}
\newcommand{\vp}{\mathbf{p}}
\newcommand{\vk}{\vec{k}}
\newcommand{\vx}{\vec{x}}
\newcommand{\om}{\omega}
\newcommand{\Om}{\Omega}
\newcommand{\ga}{\gamma}
\newcommand{\Ga}{\Gamma}
\newcommand{\gaa}{\Gamma_a}
\newcommand{\al}{\alpha}
\newcommand{\ep}{\epsilon}
\newcommand{\app}{\approx}
\newcommand{\cO}{{\cal O}}
\newcommand{\A}{{\text A}}
\newcommand{\B}{{\text B}}
\newcommand{\AB}{{\text{AB}}}
\newcommand{\nc}{\newcommand}
\nc{\beq}{\begin{equation}}
\nc{\eeq}{\end{equation}}
\nc{\beqa}{\begin{eqnarray}}
\nc{\eeqa}{\end{eqnarray}}
\def\DS {D\!\!\!\!/}
\def\A { A_\mu (x) }
\def\DM {\DS_{\, (\mu)}}
\def\O {{\cal O}}
\def\ts{\thinspace}
\def\fpi{F}
\def\gsim{\mathrel{\rlap{\lower4pt\hbox{\hskip1pt$\sim$}}
    \raise1pt\hbox{$>$}}}       

\title{Locality and Nonlinear Quantum Mechanics}
\author{Chiu Man Ho} \email{cmho@msu.edu}
   \affiliation{Department of
  Physics and Astronomy, Michigan State University, East Lansing, MI 48824, USA}
\author{Stephen~D.~H.~Hsu} \email{hsu@msu.edu}
   \affiliation{Department of
  Physics and Astronomy, Michigan State University, East Lansing, MI 48824, USA}
\date{\today}

\begin{abstract}

Nonlinear modifications of quantum mechanics generically lead to nonlocal effects which violate relativistic causality.
We study these effects using the functional Schrodinger equation for quantum fields and identify a type of nonlocality which causes nearly instantaneous entanglement of spacelike separated systems. We describe a simple example involving widely separated wave-packet (coherent) states, showing that nonlinearity in the Schrodinger evolution causes spacelike entanglement, even in free field theory.

\end{abstract}
\maketitle

\section{Introduction}

The linear structure of quantum mechanics has deep and important consequences, such as the behavior of superpositions. One is naturally led to ask whether this linearity is fundamental, or merely an approximation: Are there nonlinear terms in the Schrodinger equation?

Nonlinear quantum mechanics has been explored in \cite{NLQM,LogNonlinear,Weinberg1,Weinberg2,Haag,Kibble}. It has been observed that the {\it fictitious} violation of locality in the Einstein-Podolsky-Rosen (EPR) experiment in conventional linear quantum mechanics might become a {\it true} violation due to nonlinear effects \cite{Gisin,Polchinski} (in \cite{Polchinski} signaling between Everett branches is also discussed). This might allow superluminal communication and violate relativistic causality. These issues have subsequently been widely discussed \cite{Nonlocality,CzachorDoebner}.

Properties such as locality or causality are difficult to define in non-relativistic quantum mechanics (which often includes, for example, ``instantaneous'' potentials such as the Coulomb potential). Therefore, it is natural to adopt the framework of quantum field theory: Lorentz invariant quantum field theories are known to describe local physics with relativistic causality (influences propagate only within the light cone), making violations of these properties easier to identify. 

In this paper we are interested in {\it fundamental} nonlinearity in quantum mechanics, which is another reason for considering quantum field theory. If the evolution of quantum states is nonlinear, that should also be the case when the states in question describe quantum fields, not just individual particles.

\section{Locality and Separability}

Quantum field theory can be formulated in terms of a wavefunctional $\Psi [\phi(x),t]$ where $\phi(x)$ is a time-independent
field configuration and $t$ is the time. The functional Schrodinger equation is then given by
\beq
i \,\partial_t \,\Psi [ \phi,t ]  = \hat{H} \,\Psi [ \phi,t ] ~~.
\eeq
The Hamiltonian operator $\hat{H}$ is a sum of local operators at points $x$. For example, in scalar field theory,
\beq
\label{H}
\hat{H} = \frac{1}{2} \int d^3x ~ \left( - \frac{\delta^2}{\delta \phi^2 (x)} + \vert \nabla \phi \vert^2 + m^2 \phi^2 \right) ~~.
\eeq


Let $\Psi [ \phi,t ]  =  \psi_A [ \phi_A,t ] \,\times\, \psi_B [ \phi_B,t ] \,\times\, \cdots$, where $\phi_A (x)$ is a field configuration with support in the compact region $A$  (i.e., $\phi_A (x)$ is zero for $x \notin A$), and similarly for $\phi_B$. Assume that $A$ and $B$ are widely separated, and that the remaining factors represented by $\cdots$ do not depend on the field configuration in $A$ or $B$. This direct product structure implies, in particular, that there is no entanglement between regions $A$ and $B$. It is obviously an idealization -- in reality one might expect entanglement which decays exponentially with some correlation length such as the inverse mass gap. However, by taking $A$ and $B$ far apart we can make the approximation of no entanglement between them to be arbitrarily precise.

It is easy to show that the Schrodinger equation splits into separate equations governing $\psi_A$ and $\psi_B$:
\beq
i\, \partial_t \,\psi_A [ \phi_A,t ]  = \hat{H}_A \,\psi_A [ \phi_A,t ] ~~,
\eeq
and similarly for $B$. The subscript on the Hamiltonian $\hat{H}_A$ emphasizes that it only refers to the part of the spatial integral in (\ref{H}) over region $A$. The part of the integral over region $B$ only acts on $\psi_B$, etc.

Thus, in the absence of entanglement between $A$ and $B$, quantum mechanics in each region can be studied independently of the other. In a relativistic field theory, entanglement and other influences can propagate no faster than  the speed of light, so that if $A$ and $B$ are widely separated and initially unentangled, they will remain so for a period of time that depends on the separation.

Now, consider a nonlinear generalization of the Schrodinger equation:
\beq
i \,\partial_t\, \Psi   = \left( \hat{H} + \hat{F} ( \Psi^{\dagger} , \Psi ) \right) \Psi   ~~.
\eeq
The nonlinear term $\hat{F}$ will generically couple $\psi_A$ and $\psi_B$. Regardless of the distance between regions $A$ and $B$, the two initially unentangled states $\psi_A$ and $\psi_B$ influence each other's evolution, and typically become entangled almost instantaneously. This time evolution is illustrated in Fig. 1.

\begin{figure}[t!]
\includegraphics[height=5cm, width=4cm]{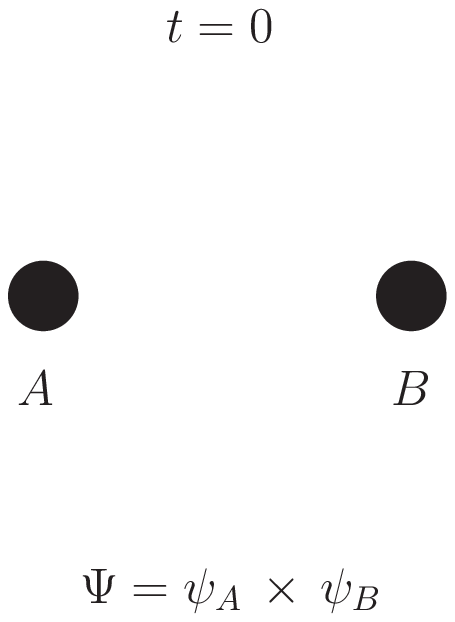}
~~~~~~~~~~~~~~~~~~~~~~~~~~~~~~~~
\includegraphics[height=5cm, width=4cm]{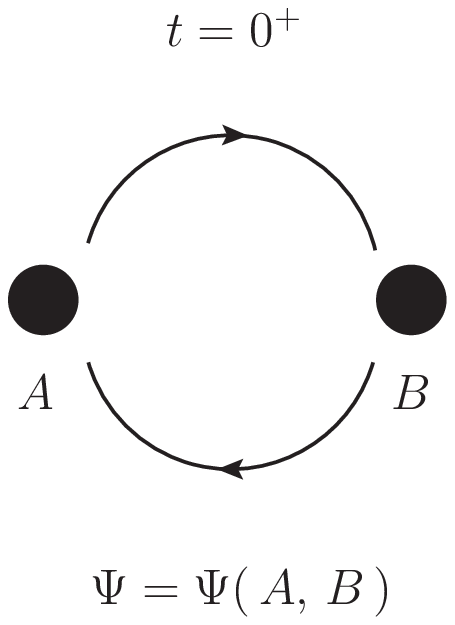}
\caption{Time evolution when nonlinearity is present. $\Psi (A,B)$ is generically an entangled state, whereas $\Psi = \psi_A \times \psi_B$ is not.}
\label{fig:self}
\end{figure}

One can also understand this from the perturbation theory point of view. Treating $\hat{F}$ as a perturbation,
we can expand $\psi_A$ and $\psi_B$ as:
\beqa
\psi_A = \psi_A^{(0)}+ \psi_A^{(1)} + \cdots ~~~~; ~~~~ \psi_B = \psi_B^{(0)}+ \psi_B^{(1)} +\cdots~~.
\eeqa
Keeping the perturbations up to the lowest order, the functional Schrodinger equation becomes
\beqa
\label{perturb}
\frac{1}{\psi_A^{(0)}}\, \left(\,i\, \partial_t -\hat{H}\,\right) \, \psi_A^{(1)} + \frac{1}{\psi_B^{(0)}}\, \left(\,i\, \partial_t -\hat{H}\,\right) \, \psi_B^{(1)}
+ \hat{F}(\psi_A^{(0)}, \psi_B^{(0)}) = 0 ~~.
\eeqa
Unless $\hat{F}(\psi_A^{(0)}, \psi_B^{(0)})$
takes the form $\hat{F}(\psi_A^{(0)}, \psi_B^{(0)}) = f_A(\psi_A^{(0)}) + f_B(\psi_B^{(0)})$,\, $\psi_A^{(1)}$ will generally be influenced by $\psi_B^{(0)}$ and vice versa. That is, the time evolution of $\psi_A$ depends on $\psi_B$ immediately at $t = 0^+$. Subsequently, a measurement of $B$ affects the state of $A$, implying entanglement. Equivalently, 
one can see that a subsequent measurement of $B$ affects the {\it probability distribution} of outcomes for $A$, implying entanglement. The manner in which $B$ affects $A$ in (\ref{perturb}) (and vice versa) is clearly nonlocal and violates relativistic causality. Similar effects do not arise in ordinary quantum field theory (i.e., assuming linear quantum dynamics and Lorentz invariance).

\section{Examples of nonlinear terms: homogeneous and otherwise}

Since quantum field theory is simply quantum mechanics of a large number of degrees of freedom (i.e., the field configurations), the discussions above apply equally well to both quantum field theory and quantum mechanics of individual particles.

The nonlinear Schrodinger equation was first considered by \cite{NLQM}, with the idea of using the nonlinearity as a possible way to resolve the difficulties associated with the quantum measurement theory. A simple example is:
\beqa
\hat{F} ( \Psi^{\dagger} , \Psi ) = \varepsilon \,|\Psi|^2 ~~,
\eeqa
which violates separability and hence locality according to our arguments above.

To maintain the separability of the wavefunction for separate systems, it was proposed in \cite{LogNonlinear} that the nonlinearity should take a logarithmic form such as
\beqa
\hat{F} ( \Psi^{\dagger} , \Psi ) = b \,\ln\,|\Psi|^2 ~~.
\eeqa
Since $\ln |\psi_A\,\psi_B|^2 = \ln |\psi_A |^2 + \ln |\psi_B|^2$, separability is maintained for an initial state which is factorizable. However, separability fails for superpositions such as identical particle states:
\beqa
\Psi = \frac{1}{\sqrt{2}}\,\left(\,\psi_A (x_1)\, \psi_B (x_2) \pm \psi_A(x_2) \, \psi_B(x_1)\,\right)~~.
\eeqa
Indeed, when the initial state $\Psi$ entangles $A$ and $B$, the log-nonlinearity causes the dynamical evolution of the $A$ system to depend on the $B$ system and vice-versa. Technically, this is somewhat different from the case of non-logarithmic interactions, where initially unentangled states become immediately entangled regardless of separation, but is nevertheless another kind of instantaneous action at a distance.

Perhaps the most systematic framework for introducing nonlinearities to quantum mechanics was provided by Weinberg \cite{Weinberg1,Weinberg2}. A key aspect of Hilbert space is that for any arbitrary complex number $Z$, the wavefunctions $\psi$ and $Z\,\psi$ represent the same physical state. It would therefore be desirable for the dynamical evolution (i.e., Schrodinger equation) to be invariant under this rescaling. Weinberg refers to this property as homogeneity. However, all of the proposals for nonlinear quantum mechanics suggested by \cite{NLQM,LogNonlinear} lack this property. In contrast, Weinberg's framework respects the homogeneity condition and Galilean invariance explicitly. The nonlinear Schrodinger equations proposed by \cite{Haag} and \cite{Kibble} also satisfy the homogeneity condition, but \cite{Haag} requires an arbitrary vector potential and \cite{Kibble} violates Galilean invariance.

A simple example satisfying the homogeneity condition is:
\beqa
\hat{F} ( \Psi^{\dagger} , \Psi ) = \frac{ \Psi^{\dagger}\,\hat{O}_1\,\Psi}{\Psi^{\dagger}\,\hat{O}_2\,\Psi} ~~,
\eeqa
where $\hat{O}_1$ and $\hat{O}_2$ are some Hermitian operators. In general, this leads to non-separability and hence nonlocality. An exception is when both $\hat{O}_1$ and $\hat{O}_2$ commute with $\hat{H}$. In this case, the nonlinear terms only cause constant shifts to the Hamiltonian.

One possibility discussed by Weinberg \cite{Weinberg2} and by Polchinski \cite{Polchinski} occurs if the denominator in $\hat{F}$ is the magnitude squared of the entire wavefunction. If the branch of the wavefunction occupied by the observer is only a small component of the total (i.e., this might be the case after many decoherent outcomes are recorded in the memory of that observer, assuming of course that decoherence continues to operate as usual in the presence of nonlinearity), then the effect of the nonlinear term is suppressed for that observer, even if at the fundamental level quantum mechanics is highly nonlinear. In this scenario we do not expect to observe any nonlinearity so late in the history of the universe.

\section{Example: free field theory}

As a specific example, we can consider nonlinear quantum dynamics of free field theory, where the Hamiltonian is diagonal in the Fock basis, and the nonlinear term can be calculated explicitly. The absence of interactions also eliminates any subtleties associated with renormalization, at least in the case of linear quantum dynamics. We will find that the properties of state $\psi_A$ influence physics in region $B$ and vice-versa, regardless of the distance between the two regions.

Let $\psi_A^{(0)}$ and $\psi_B^{(0)}$ be coherent state wavefunctionals, so the states $A$ and $B$ are semiclassical (minimum uncertainty) configurations such as wave packets, localized in regions $A$ and $B$ respectively. (This localization need not be exact; it may hold to exponential accuracy.)
\beqa
\psi_A^{(0)} [ \phi_A ] = \langle \phi_A \vert a_A \rangle = e^{\Omega[\,a_A,\,\phi_A \,]} ~~~~; ~~~~ \psi_B^{(0)} [ \phi_B ] = \langle \phi_B \vert a_B \rangle = e^{\Omega[\,a_B,\,\phi_B \,]}~~,
\eeqa
where $\langle \phi \vert$ is an eigenstate of the field operator $\hat{\phi}$ and the coherent state $| a \rangle$ is an eigenstate of the annihilation operator $\hat{a}_{\mathbf{k}}$:\, $\hat{a}_{\mathbf{k}} | a \rangle =  a( \mathbf{k} ) | a \rangle $. More explicitly, $| a \rangle$
has the form
\beqa
\label{creation}
| a \rangle =  \exp \left( \,\,\int d^3\mathbf{k} \,\, a(\mathbf{k})\, \hat{a}_{\mathbf{k}}^{\dagger} \,\,\right) \, | 0 \rangle ~~,
%
%
\eeqa
and, for $a(\mathbf{k}) = a_{A,B} (\mathbf{k})$, can be interpreted directly as particle states localized in the $A,B$ regions.

In our notation, $\hat{a}_{\mathbf{k}}$ is an (annihilation) operator but $a( \mathbf{k} )$ is a complex function of the momentum $\mathbf{k}$. Each of $a_A$ and $a_B$ is specified by their Fourier transforms $a_{A,B} ( \mathbf{k} )$. The full expression for $\Omega[a,\phi]$ can be found in \cite{Hsu}:
\beqa
\label{Omega}
\Omega[a,\phi] &=& -\frac12\,\int\, d^3 \mathbf{k}\,\, a ({\mathbf{k}}) \,a (-\mathbf{k})\, e^{2\,i\,\omega_{\mathbf{k}}\,t}
-\frac12\,\int\, d^3 \mathbf{k}\,\,\omega_{\mathbf{k}}\, \phi(\mathbf{k})\,\phi(-\mathbf{k})  \nonumber \\
&& + \int\, d^3 \mathbf{k}\,\,\sqrt{2\,\omega_{\mathbf{k}}}\,\, a( \mathbf{k})\, e^{i\,\omega_{\mathbf{k}}\,t}\, \phi(-\mathbf{k})~~.
\eeqa
If $a_A( \mathbf{x} )$ has its support in region $A$, and similarly with $B$, then approximate factorization for widely-separated coherent states holds: $\Psi [ \phi,t ]  \approx  \psi_A [ \phi_A,t ] \,\times\, \psi_B [ \phi_B,t ] \,\times\, \cdots~$. Details of the proof are given in the Appendix. 
Factorization may not be an exact property, but holds to exponential accuracy as the regions $A$ and $B$ become widely separated. Particles cannot be {\it completely} localized in $A,B$ without invoking non-analytic functions $a_{A,B} (\mathbf{x})$.

Then $\hat{F}$ is a time-dependent function of the two coherent states localized at $A$ and $B$. To focus on a reduced subset of the degrees of freedom, consider rescalings: $a_A \rightarrow \alpha \,a_A$ and $a_B \rightarrow \beta\, a_B$.
To be definite, we can take $\hat{F} \propto |\,\Psi_A \Psi_B \,|$ to some power. IF factorization continues to hold, the Schrodinger equation (\ref{perturb}) has the form
\beqa
\label{gh}
g_A(\alpha) + g_B (\beta) = h_A (\alpha) \, h_B(\beta)~~,
\eeqa
where $\{\,g_A, \,h_A\,\}$ and $\{\,g_B, \,h_B\,\}$ are functions depending only on $\alpha$ and $\beta$ respectively. This condition cannot hold for all choices of $\alpha$ and $\beta$. Therefore, separability is violated for at least some states and as a consequence we have nonlocality. The conclusion is the same for any form of $\hat{F}$ except a logarithm, which has the problems discussed previously.

To summarize, we have verified that nonlinear modifications to the evolution of the wavefunctional $\Psi [ \phi , t ]$ lead to nonlocality even for free field theory (e.g., a non-interacting scalar, or photons in the absence of charged particles). This demonstration is not subject to renormalization or related subtleties. We can describe our setup in physical terms. At $t=0$ we have a scalar particle state localized in region $A$, and another in region $B$. (These are described by the states $| a_A \rangle$ and $| a_B \rangle$; see (\ref{creation}).) They are completely unentangled. The steps leading to (\ref{gh}) show that nonlinearity in the functional Schrodinger equation generically leads to entanglement of these previously unentangled, and spacelike separated, particles. This could of course be recast in terms of an ordinary quantum mechanics description of the wavefunctions of the individual particles at $A$ and $B$.

\section{Conclusions and Discussions}

Our results suggest that nonlinearity in quantum mechanics leads to violation of relativistic causality. We gave a formulation in terms of approximately factorized (unentangled) wavefunctions describing spacelike separated systems. Nonlinearity creates almost instantaneous entanglement of the two systems, no matter how far apart. Perhaps our results are related to what Weinberg \cite{Weinberg3} meant when he wrote ``... I could not find any way to extend the nonlinear version of quantum mechanics to theories based on Einstein's special theory of relativity ... At least for the present I have given up on the problem: I simply do not know how to change quantum mechanics by a small amount without wrecking it altogether.''

Finally, it may be interesting to consider nonlinear modification of the Wheeler-DeWitt equation (i.e., the Schrodinger equation for geometries in quantum gravity). Because there is no intrinsic notion of locality in quantum gravity, nonlinear modifications might not lead to catastrophic consequences. However, it seems likely that the nonlinearities would find their way into quantum mechanics on semiclassical spacetimes, as we have considered here. In that case, there would be unwelcome violations of locality.

\section{Acknowledgements}

This work was supported by the Office of the Vice-President for Research and Graduate Studies at Michigan State University.


\section{Appendix:  ~Proof of Approximate Factorization}

In this appendix, we will first provide the proof for $\Psi [\, \phi\, ]
\approx \psi_A [\, \phi_A\, ] \times \psi_B [\, \phi_B\, ]$ and then generalize it to
$\Psi [\, \phi\, ] \approx \psi_A [\, \phi_A\, ] \times \psi_B [\, \phi_B\, ] \times \cdots$.

To begin with, we define coherent wave packet states (see, for instance, the discussions below Eq.(3.57) in \cite{Itzykson}) with disjoint support in two widely-separated regions $A$ and $B$. These states are constructed using the ordinary creation operators. Let $a(\mathbf{k}) = a_A (\mathbf{k}) + a_B (\mathbf{k})$. In coordinate space, $a_A (\mathbf{x})$ and $a_B (\mathbf{x})$ are {\it defined} to have support only in regions $A$ and $B$ respectively. Let
\beqa
| a \rangle &=&  \exp \left( \,\,\int d^3\mathbf{k} \,\, \left(\,a_A (\mathbf{k}) + a_B (\mathbf{k})\,\right)\, \hat{a}_{\mathbf{k}}^{\dagger} \,\,\right) \, | 0 \rangle  \nonumber \\ &=& \exp \left( \, \int d^3\mathbf{k} \, a_A (\mathbf{k}) \, \hat{a}_{\mathbf{k}}^{\dagger} \right) ~   \exp  \left(  \, \int d^3\mathbf{k} \,  a_B (\mathbf{k})  \, \hat{a}_{\mathbf{k}}^{\dagger}   \right)   ~ | 0 \rangle                                         \, ,
\eeqa
which describes a wave packet in region $A$ and another in region $B$. These wave packets are created using the usual creation
operator $\hat{a}_{\mathbf{k}}^{\dagger}$.

Now consider
\beqa
\Psi [\, \phi\, ] = \langle \phi  \vert  a \rangle = \exp  \Omega[\,a,\,\phi \,] \,.
\eeqa
For convenience (we can relax this assumption later), let $\phi(\mathbf{k}) = \phi_A (\mathbf{k}) + \phi_B (\mathbf{k})$, where, in coordinate space, $\phi_A (\mathbf{x})$ and $\phi_B (\mathbf{x})$ have support only in regions $A$ and $B$ respectively.

According to (\ref{Omega}) in the paper, we have
\beqa
\label{OmegaApp}
\Omega[a,\phi] &=& -\frac12\,\int\, d^3 \mathbf{k}\,\, a ({\mathbf{k}}) \,a (-\mathbf{k})\, e^{2\,i\,\omega_{\mathbf{k}}\,t}
-\frac12\,\int\, d^3 \mathbf{k}\,\,\omega_{\mathbf{k}}\, \phi(\mathbf{k})\,\phi(-\mathbf{k})  \nonumber \\
&& + \int\, d^3 \mathbf{k}\,\,\sqrt{2\,\omega_{\mathbf{k}}}\,\, a( \mathbf{k})\, e^{i\,\omega_{\mathbf{k}}\,t}\, \phi(-\mathbf{k})\,,
\eeqa
where, for simplicity, we have suppressed the time-dependent factors like $e^{i\,\omega_{\mathbf{k}}\,t}$ in the original equation
(or one can think of this as the case with $t=0$). Now, study each of the above integrals separately:

1. Consider
\beqa
I_1 &\equiv&  -\frac12\,\int\, d^3 \mathbf{k}\,\, a ({\mathbf{k}}) \,a (-\mathbf{k}) \nonumber \\
&=& -\frac12\,\int\, d^3 \mathbf{k}\,\,\left[\, a_A ({\mathbf{k}}) \,a_A (-\mathbf{k})+ a_B ({\mathbf{k}}) \,a_B (-\mathbf{k})
+ a_A ({\mathbf{k}}) \,a_B (-\mathbf{k})+a_B ({\mathbf{k}}) \,a_A (-\mathbf{k}) \, \right]\,. \nonumber \\
\eeqa
The first two terms in $I_1$ are the normalization factors. The third and the fourth terms are zero for the following reason:
\beqa
\int\, d^3 \mathbf{x}\,\, a_A ({\mathbf{x}}) \,a_B (\mathbf{x}) &=& \int\, d^3 \mathbf{x}\,\,\int\, d^3 \mathbf{k}\,\,\int\, d^3 \mathbf{p}\,\,
a_A ({\mathbf{k}}) \,e^{i\,\mathbf{k}\cdot \mathbf{x}}\,\,a_B (\mathbf{p}) \,e^{i\,\mathbf{p}\cdot \mathbf{x}} \nonumber \\
&=& \int\, d^3 \mathbf{k}\,\,\int\, d^3 \mathbf{p}\,\,
a_A ({\mathbf{k}})\,a_B (\mathbf{p}) \,\delta^{(3)}(\mathbf{k}+\mathbf{p}) \nonumber \\
&=& \int\, d^3 \mathbf{k}\,\,a_A ({\mathbf{k}})\,a_B (-\mathbf{k})\,.
\eeqa
Since $a_A (\mathbf{x})$ and $a_B (\mathbf{x})$ have support only in regions $A$ and $B$ respectively, we have
$\int\, d^3 \mathbf{x}\,\, a_A ({\mathbf{x}}) \,a_B (\mathbf{x}) = 0$ and hence
$\int\, d^3 \mathbf{k}\,\,a_A ({\mathbf{k}})\,a_B (-\mathbf{k}) =0$. Similarly,
$\int\, d^3 \mathbf{k}\,\,a_B ({\mathbf{k}})\,a_A (-\mathbf{k}) =0$. As a result,
\beqa
I_1 = -\frac12\,\int\, d^3 \mathbf{k}\,\, a_A ({\mathbf{k}}) \,a_A (-\mathbf{k})
-\frac12\,\int\, d^3 \mathbf{k}\,\, a_B ({\mathbf{k}}) \,a_B (-\mathbf{k})\,.
\eeqa

2. Consider
\beqa
I_2 &\equiv&  -\frac12\,\int\, d^3 \mathbf{k}\,\, \omega_{\mathbf{k}}\,\phi ({\mathbf{k}}) \,\phi (-\mathbf{k}) \nonumber \\
&=& -\frac12\,\int\, d^3 \mathbf{k}\,\,\omega_{\mathbf{k}}\,\left[\, \phi_A ({\mathbf{k}}) \,\phi_A (-\mathbf{k})+ \phi_B ({\mathbf{k}}) \,\phi_B (-\mathbf{k})
+ \phi_A ({\mathbf{k}}) \,\phi_B (-\mathbf{k})+\phi_B ({\mathbf{k}}) \,\phi_A (-\mathbf{k}) \, \right]\,. \nonumber \\
\eeqa
Again, the first two terms in $I_2$ are the normalization factors. The third and the fourth terms approach zero for the following reasons.
First of all, we have $\phi_A ({\mathbf{k}}) = \int\, d^3 \mathbf{x}\,\, \phi_A ({\mathbf{x}})\,e^{-i\,\mathbf{k}\cdot \mathbf{x}}$.
Suppose that region $B$ is at a distance $d$ from region $A$, and (for simplicity; this is not essential) that the shapes of the functions are the same in their respective domains: $\phi_{B} (\mathbf{x}) = \phi_{A} ({\mathbf{x}}+{\mathbf{d}})$. Then it follows that 
\beqa
\phi_B ({\mathbf{k}}) &=& \int\, d^3 \mathbf{x}\,\, \phi_A ({\mathbf{x}}+{\mathbf{d}})\,e^{-i\,\mathbf{k}\cdot \mathbf{x}}
~~~~;~~~~ |{\mathbf{d}}|=d \nonumber \\
&=& e^{i\,\mathbf{k}\cdot \mathbf{d}} \,\int\, d^3 \mathbf{y}\,\, \phi_A ({\mathbf{y}})\,e^{-i\,\mathbf{k}\cdot \mathbf{y}} \nonumber \\
&=& e^{i\,\mathbf{k}\cdot \mathbf{d}}\, \phi_A ({\mathbf{k}})\,.
\eeqa
(If the functions $\phi_{A,B} (\mathbf{x})$ differ in their respective domains, we cannot express $\phi_B ({\mathbf{k}})$ in terms of $\phi_A ({\mathbf{k}})$, but the conclusions below still hold; the main point is that the two Fourier transforms differ by a rapidly varying phase factor.) Consequently,
\beqa
\int\, d^3 \mathbf{k}\,\, \omega_{\mathbf{k}}\,\phi_A ({\mathbf{k}}) \,\phi_B (-\mathbf{k}) 
= \int\, d^3 \mathbf{k}\,\, \omega_{\mathbf{k}}\, \phi_A ({\mathbf{k}}) \,\phi_A (-\mathbf{k})\, e^{-i\,\mathbf{k}\cdot \mathbf{d}}\,.
\eeqa
Note that the integral over $|\,\omega_{\mathbf{k}}\, \phi_A ({\mathbf{k}}) \,\phi_A (-\mathbf{k})\,|$ must exist for reasonable configurations.
For widely-separated regions $A$ and $B$ with $d\rightarrow \infty$, we can apply the Riemann-Lebesgue lemma to conclude that
$\int\, d^3 \mathbf{k}\,\, \omega_{\mathbf{k}}\, \phi_A ({\mathbf{k}}) \,\phi_A (-\mathbf{k})\, 
e^{-i\,\mathbf{k}\cdot \mathbf{d}} \,\rightarrow \,0$ (exponentially in $d$). Thus, $\int\, d^3 \, \mathbf{k}\,\,\omega_{\mathbf{k}}\,\phi_A ({\mathbf{k}})\,\phi_B (-\mathbf{k}) \,\rightarrow \, 0$. Similarly,
$\int\, d^3\, \mathbf{k}\,\,\omega_{\mathbf{k}}\,\phi_B ({\mathbf{k}})\,\phi_A (-\mathbf{k}) \,\rightarrow \, 0$. As a result,
\beqa
I_2 \approx -\frac12\,\int\, d^3 \mathbf{k}\,\, \omega_{\mathbf{k}}\, \phi_A ({\mathbf{k}}) \,\phi_A (-\mathbf{k})
-\frac12\,\int\, d^3 \mathbf{k}\,\, \omega_{\mathbf{k}}\, \phi_B ({\mathbf{k}}) \,\phi_B (-\mathbf{k})\,.
\eeqa

3. Consider
\beqa
I_3 &\equiv& \int\, d^3 \mathbf{k}\,\, \sqrt{2\,\omega_{\mathbf{k}}}\,a ({\mathbf{k}}) \,\phi (-\mathbf{k}) \nonumber \\
&=& \int\, d^3 \mathbf{k}\,\,\sqrt{2\,\omega_{\mathbf{k}}}\,\left(\, a_A ({\mathbf{k}}) \,\phi_A (-\mathbf{k})+ a_B ({\mathbf{k}}) \,\phi_B (-\mathbf{k}) + a_A ({\mathbf{k}}) \,\phi_B (-\mathbf{k})+a_B ({\mathbf{k}}) \,\phi_A (-\mathbf{k}) \, \right)\,. \nonumber \\
\eeqa
Following similar arguments as above, one can show that the mixed terms vanish exponentially with the separation $d$:
$\int\, d^3 \, \mathbf{k}\,\,\sqrt{2\,\omega_{\mathbf{k}}}\, a_{\{A,B\}} ({\mathbf{k}})\,\phi_{\{B,A\}} (-\mathbf{k}) \,\rightarrow \,0$. 
As a result,
\beqa
I_3 \approx \int\, d^3 \mathbf{k}\,\, \sqrt{2\,\omega_{\mathbf{k}}}\, a_A ({\mathbf{k}}) \,\phi_A (-\mathbf{k})
+\int\, d^3 \mathbf{k}\,\, \sqrt{2\,\omega_{\mathbf{k}}}\, a_B ({\mathbf{k}}) \,\phi_B (-\mathbf{k})\,.
\eeqa

Therefore, we have just shown that
\beqa
\Omega[a,\phi] \approx \Omega[a_A,\phi_A] + \Omega[a_B,\phi_B]\,,
\eeqa
where
\beqa
\Omega[a_A,\phi_A] &=& -\frac12\,\int\, d^3 \mathbf{k}\,\, a_A ({\mathbf{k}}) \,a_A (-\mathbf{k})
-\frac12\,\int\, d^3 \mathbf{k}\,\,\omega_{\mathbf{k}}\, \phi_A(\mathbf{k})\,\phi_A(-\mathbf{k})  \nonumber \\
&& + \int\, d^3 \mathbf{k}\,\,\sqrt{2\,\omega_{\mathbf{k}}}\,\, a_A( \mathbf{k})\, \phi_A(-\mathbf{k})\,, \\
\Omega[a_B,\phi_B] &=& -\frac12\,\int\, d^3 \mathbf{k}\,\, a_B ({\mathbf{k}}) \,a_B (-\mathbf{k})
-\frac12\,\int\, d^3 \mathbf{k}\,\,\omega_{\mathbf{k}}\, \phi_B(\mathbf{k})\,\phi_B(-\mathbf{k})  \nonumber \\
&& + \int\, d^3 \mathbf{k}\,\,\sqrt{2\,\omega_{\mathbf{k}}}\,\, a_B( \mathbf{k})\, \phi_B(-\mathbf{k})\,.
\eeqa
We thus conclude that
\beqa
\Psi [\, \phi\, ] &=& \langle \phi_ \vert a \rangle = e^{\Omega[\,a,\,\phi \,]} \nonumber \\
&\approx& e^{\Omega[\,a_A,\,\phi_A \,]} \times e^{\Omega[\,a_B,\,\phi_B \,]} \nonumber \\
&=& \psi_A [\, \phi_A\, ] \times \psi_B [\, \phi_B\, ]\,.
\eeqa

So far, our proof has focused on two regions $A$ and $B$, but it is clear that it can be easily generalized to any number of widely separated regions. In the most general case, we obtain 
\beqa
\Omega[a,\phi] \approx \Omega[a_A,\phi_A] + \Omega[a_B,\phi_B] + \cdots\,\,,
\eeqa
where $``\cdots"$ denotes the dependence of $\Psi$ on any regions outside $A$ and $B$. Consequently, we will have
\beqa
\Psi [\, \phi\, ]
\approx \psi_A [\, \phi_A\, ] \times \psi_B [\, \phi_B\, ] \times \cdots \,\,.
\eeqa




\begin{thebibliography}{99}

\bibitem{NLQM}
  L. de Broglie, ``Non-Linear Wave Mechanics -- A Causal Interpretation", Elsevier, Amsterdam, 1950;
  P.~M.~Pearle,
  Phys.\ Rev.\ D {\bf 13}, 857 (1976);
  B.~Mielnik,
  Commun.\ Math.\ Phys.\  {\bf 37}, 221 (1974).

\bibitem{LogNonlinear}
  I.~Bialynicki-Birula and J.~Mycielski,
  Annals Phys.\  {\bf 100}, 62 (1976).

\bibitem{Weinberg1}
  S.~Weinberg,
  Phys.\ Rev.\ Lett.\  {\bf 62}, 485 (1989).

\bibitem{Weinberg2}
  S.~Weinberg,
  Annals Phys.\  {\bf 194}, 336 (1989).

\bibitem{Haag}
  R.~Haag and U.~Bannier,
  Commun.\ Math.\ Phys.\  {\bf 60}, 1 (1978).

\bibitem{Kibble}
  T.~W.~B.~Kibble,
  Commun.\ Math.\ Phys.\  {\bf 64}, 73 (1978).

\bibitem{Gisin}
  N.~Gisin,
  Helv.\ Phys.\ Acta {\bf 62}, 363 (1989);
  N.~Gisin, Phys.\ Rev.\ A {\bf 143}, 1 (1990);
  M. Czachor, Found. Phys. Lett. {\bf 4}, 351 (1991).

\bibitem{Polchinski}
  J.~Polchinski,
  Phys.\ Rev.\ Lett.\  {\bf 66}, 397 (1991).

\bibitem{Nonlocality}
  G.~Svetlichny,
  Found.\ Phys.\ {\bf 28}, 131 (1998);
  G.~Svetlichny,
  Int.\ J.\ Theor.\ Phys.\  {\bf 44}, 2051 (2005);
  C.~Simon, V.~Bu\v{z}ek and N.~Gisin,
  Phys.\ Rev.\ Lett.\  {\bf 87}, 170405 (2001);
  B.~Mielnik,
  quant-ph/0012041;
   W.~Luecke,
  quant-ph/9904016;
  H.~D.~Doebner,
  quant-ph/9803011;
  A.~Caticha,
  Phys.\ Lett.\ A {\bf 244}, 13 (1998);
  W.~Puszkarz,
  quant-ph/9710010;
  W.~Puszkarz,
  quant-ph/9903010;
  W.~Puszkarz,
  quant-ph/9905046;
  H.~-T.~Elze,
  Int.\ J.\ Theor.\ Phys.\  {\bf 47}, 455 (2008);

\bibitem{CzachorDoebner}
  Marek Czachor and H. -D. Doebner,
  Phys. Lett. A {\bf 301}, 139 (2002). These authors suggest a modified definition of measurement
  (appropriate for nonlinear quantum mechanics) that avoids nonlocality of the Gisin-Polchinski type
  (Refs. \cite{Gisin,Polchinski}), but it does not address the question of separability that is our main
  interest in this paper.





\bibitem{Hsu}
  S.~D.~H.~Hsu,
  Phys.\ Lett.\ B {\bf 555}, 92 (2003);
  T.~M.~Gould, S.~D.~H.~Hsu and E.~R.~Poppitz,
  Nucl.\ Phys.\ B {\bf 437}, 83 (1995).



\bibitem{Weinberg3} S. Weinberg, {\it Dreams of a Final Theory}, Hutchison (1993).

\bibitem{Itzykson} C. Itzykson and J.-B. Zuber, {\it Quantum Field Theory}, Dover (2006).

\end{thebibliography}
\end{document}